\documentclass[twocolumn, preprintnumbers, noshowkeys, noshowpacs floatfix]{revtex4}
\usepackage{graphicx}
\usepackage[english]{babel}
\newcommand{\degr}{$^{\circ}$}

\begin{document}

\preprint{To appear in {\em Review of Scientific Instruments}, 2005}

\title{Spectroscopy with Multichannel Correlation Radiometers}

\author{A.I.~Harris}
\affiliation{Department of Astronomy, University of Maryland, \\
  College Park, MD 20742}
\email{harris@astro.umd.edu}

\begin{abstract}

Correlation radiometers make true differential measurements in power
with high accuracy and small systematic errors.  This receiver
architecture has been used in radio astronomy for measurements of
continuum radiation for over 50 years; this article examines
spectroscopy over broad bandwidths using correlation techniques.
After general discussions of correlation and the choice of hybrid
phase experimental results from tests with a simple laboratory
multi-channel correlation radiometer are shown.  Analysis of the
effect of the input hybrid's phase shows that a 90\degr\ hybrid is
likely to be the best general choice for radio astronomy, depending on
its amplitude match and phase flatness with frequency.  The laboratory
results verify that the combination of the correlation architecture
and an analog lag correlator is an excellent method for spectroscopy
over very wide bandwidths.

\end{abstract}



\maketitle

\section{Introduction}

Modern analog lag correlators are capable of autocorrelation
spectroscopy over wide bandwidths \cite{hz01}.  This article examines
an application of analog lag correlators, that of measuring
cross-correlation functions for spectroscopy with correlation
detection techniques.  Although variants of the correlation detection
schemes are common for radio continuum radiometry, it seems that no
one has yet adapted the architecture for spectroscopy (multichannel
radiometry) for high-resolution spectroscopy with a single telescope.
A cross-correlation spectrometer on a single-aperture instrument would
have the same excellent stability as cross-correlation spectrometers
in aperture synthesis arrays.  In addition, as shown in
Section~\ref{sec:anaxc}, a multichannel correlation radiometer can
share one analog cross-correlation backend spectrometer between two
sky positions, providing a dual-beam system with about half the
backend spectrometer cost and complexity of a receiver with dual
total-power spectrometers.

Correlation radiometers have made accurate radio and millimeter-wave
continuum intensity measurements of the radio sky, the Cosmic
Microwave Background, of rapidly changing scenes, and polarimetry
(e.g.\ \cite{haslam74, jarosik03a, predmore85, rholfs00_crx,
koistinen02}).  A number of authors have described specific
architectures and examined the operation and sensitivity of
correlation radiometers in absolute terms and their suppression of
effects from the 1/$f$ noise common to amplifiers \cite{fano51, gs55a,
blum59, tiuri64, colvin61, faris67, predmore85, seiffert02,
mennella03}.  In the following, Section~\ref{sec:corrdet} contains a
general discussion of correlation and Section~\ref{sec:hybrid} gives
an analysis of the choice of hybrid phase, information that is not
readily available elsewhere.  Section~\ref{sec:anaxc} shows
experimental results from tests with a simple laboratory multichannel
correlation radiometer that verify the theoretical expectations.

\section{Correlation detection \label{sec:corrdet}}

Ryle introduced correlation detection to radio astronomy with his
invention of the phase-switching interferometer \cite{ryle52}.  Ryle's
interferometer squared the sum of the voltages from two antennas after
modulating the phase of one antenna's signal.  Maintaining phase
sensitivity by multiplying voltages instead of detecting the total
power alone made this a correlating instead of a phased array of
antennas.  Phase switching the signal from one antenna was the key
element in the method's success, as it separated the desired
cross-product of the two antenna's voltages from total power signals
from the individual antennas.  Correlation detection with rapid phase
switching brought a substantial improvement in instrumental stability
since gain and noise fluctuations of amplifiers on the two antennas
were uncorrelated in time; the only correlated signal, that common to
the two antennas, was from the astronomical source.  Communication
engineers \cite{lee49, lee50} had already recognized that correlation
techniques were valuable for retrieving small periodic signals in
noise, and eventually came to view cross-correlation as a method of
producing an optimal filter: it selects the component of the input
signal at one multiplier input with waveform equal to the reference
signal at the other multiplier input \cite{pzpeebles93_corr}.  Viewed
in this light, synchronous detection is a familiar example of a
correlation receiver.
  
\begin{figure}
\includegraphics[width=3.2in]{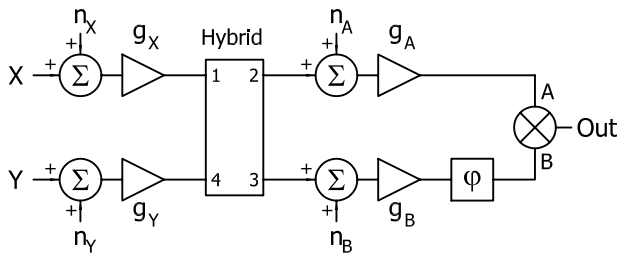} 
\caption{General model for a continuous comparison radiometer.  The
central block represents the input hybrid, with input ports numbered 1
and 4 and outputs 2 and 3. General noise and gain terms $n$ and $g$
and a phase shift $\varphi$ complete the model. \label{fig:blkcrx}}
\end{figure}

Adding a four-port circuit (a ``hybrid'') to correlation detection
combines signals from two regions of the focal plane of a single
aperture telescope and redistributes them before amplification.  This
allows the correlation technique to be used for single-dish
observations.  Figure~\ref{fig:blkcrx} shows the signal flow through a
correlation radiometer; the components to the right of the hybrid are
equivalent to a spatial interferometer's signal path.  The terminology
for correlation detection is unfortunately muddled.  Communications
engineers use the term correlation receiver to describe what radio
astronomers usually think of as a spatial interferometer or a
synchronous detector, while single-dish astronomical instruments take
the name ``correlation receiver'' \cite{colvin61, tiuri64} or the more
apt ``continuous comparison receiver'' \cite{blum59, predmore85}.
More recently, the same architecture has been dubbed the
``pseudo-correlation receiver,'' apparently based on a detail of the
multiplier implementation.

Continuous comparison detection with a single-aperture telescope is
the complement to the two-element spatial interferometer: an
interferometer is sensitive to the correlated signals from two
different regions of an aperture plane, while the continuous
comparison radiometer extracts the uncorrelated part of the signals
from two different regions of a focal plane.  The input hybrid
combines voltages from the source and reference positions in the focal
plane, $v_s$ and $v_r$, with different but known phase shifts before
amplification.  Cross-correlating the signals from the two amplifiers
with the proper phase shift extracts the power difference between the
two positions, $v_{out} \propto G \left( \langle v_s^2 \rangle -
\langle v_r^2 \rangle \right)$, where the angle brackets denote an
average over a time long compared with the reciprocal of the input
bandwidth.  The single-aperture continuous comparison radiometer has
signal paths that are as similar as possible, so subsequent
amplification and processing operate equally on the signals from both
inputs.  An amplifier gain fluctuation, for instance, has exactly the
same effect on the signals from both positions in the focal plane.
Correlated terms, including amplifier noise, average away with time as
$1/\sqrt{B \tau}$, where $B$ is the predetection bandwidth and $\tau$
is the integration time.  With a differential measurement, gain
fluctuations have no large noise term to amplify, greatly reducing the
excess noise across the spectrum.  Excess noise is a particular
problem for instruments with wide bandwidths because the intrinsic
measurement noise is proportional to $1/\sqrt{B}$ by the radiometer
equation \cite{dicke46}.  Fluctuations add noise in individual
channels and instrumental structure across the spectrum.

A true differential measurement can greatly improve the stability of a
radiometer compared with conventional total power measurements.  Power
amplification in a typical heterodyne radiometer is about $10^{12}$,
so it is not surprising that the most common limit to the stability of
radiometric measurements is electronic and optical gain instability in
time.  Consider a total power radiometer with gain $G$, input source
voltage $v_s$, and system noise voltage $v_n$.  Its detector output
voltage $v_{out}$ is proportional to $G\langle v_{detector}^2
\rangle$, or $v_{out} \propto G \left( \langle v_s^2 \rangle + \langle
v_n^2\rangle \right)$, plus the small uncorrelated cross term
$2\langle v_s v_n \rangle$ that averages away with time as $1/\sqrt{B
\tau}$.  Spatial chopping and differencing between source and
reference positions on the sky largely eliminates the relatively large
noise signal on average, but even tiny fluctuations in system gain $G$
or noise power $\langle v_n^2 \rangle$ at the chop frequency can
easily be much larger than the weak signal, $\Delta \!\!  \left( G
\langle v_n^2 \rangle \right) \gg \langle v_s^2 \rangle$, and can
dominate the integrated signal.  In contrast with Dicke's
\cite{dicke46} scheme of sequential switching between astronomical and
reference signals with a total power radiometer, the continuous
comparison technique's simultaneous treatment of signal and reference
positions provides a true differential measurement that greatly
reduces the effects of time-variable gain fluctuations.  Differencing
without mechanically changing the optical system can also help reduce
instabilities induced effects from microphonics or changing
standing-wave structure that affect some types of receivers (e.g.\
local oscillator power modulation from a focal plane chopper or
nutating secondary) and other low-level effects.

\section{Choice of hybrid \label{sec:hybrid}}

Figure~\ref{fig:blkcrx} defines the noise and gain variables for the
following system analysis.  Generalized noise voltages $n_{X,Y}$ and
voltage gains $g_{X,Y}$ affect the voltages $x$ and $y$ at inputs X
and Y.  The hybrid's output is a phase-shifted mixture of its input
voltages.  Further equivalent noise voltages $n_{A,B}$ and voltage
gains $g_{A,B}$ follow at each output.  Noise voltages from components
following the hybrid will be in phase and in quadrature (denoted I and
Q) with the signal phase, so the noise terms are $n = n_{I}/\sqrt{2} +
jn_{Q}/\sqrt{2}$, where $j = \sqrt{-1}$ and the total noise power is
$\langle n^2 \rangle = \langle n_{I}^2 \rangle + \langle n_{Q}^2
\rangle$.  Solving for all noise and gain components in
Fig.~\ref{fig:blkcrx} is too messy to clearly show the circuit's
properties.  It is clearer to solve two basic cases, one with all gain
and noise following an ideal hybrid and one with all preceding it.
Nonideal effects are straightforward to include as modifications to
these ideal cases.

Most astronomical continuous comparison radiometers incorporate a
waveguide magic tee 180\degr\ hybrid \cite{jarosik03a, seiffert02},
although at least one has used a 90\degr\ quasi-optical hybrid
\cite{predmore85}, and good branch-line 90\degr\ hybrids now exist at
frequencies to hundreds of gigahertz \cite{srikanth01, kooi04}.  The
correlator's output has significantly different behavior for the
180\degr\ and 90\degr\ hybrids.  A lossless hybrid's scattering matrix
relates its output voltages to its input voltages, with ports numbered
as in Figure~\ref{fig:blkcrx}, as:
\begin{equation}
\left[ 
\begin{array}{c}
v_{o1} \\  v_{o2} \\ v_{o3} \\ v_{o4} 
\end{array}
\right]
 = -j
\left[
\begin{array}{cccc}
0 & \beta & \alpha \,e^{j\theta} & 0 \\
\beta & 0 & 0 & -\alpha\,e^{-j\theta} \\
\alpha\,e^{j\theta} & 0 & 0 & \beta \\
0 & -\alpha \,e^{-j\theta} & \beta & 0
\end{array} 
\right]
\left[ 
\begin{array}{c}
v_{i1} \\  v_{i2} \\ v_{i3} \\ v_{i4} 
\end{array}
\right]
 \; .
\label{eq:smatrix}
\end{equation}
Here $\alpha$ and $\beta$ are voltage coupling coefficients with
$\alpha^2 + \beta^2 = 1$.  Zeros along the diagonal indicate that the
ports are perfectly matched, and zeros on the cross-diagonal indicate
no coupling between isolated ports.  (A lossless hybrid must have
these terms equal to zero to satisfy the unitary property of a
lossless network $S$ matrix \cite{pozar98_spar}.)  The phase angle
$\theta$ in Eq.~(\ref{eq:smatrix}) may vary arbitrarily in theory,
with $\theta = 0$ and $\pi/2$ corresponding to realizable devices: the
fully asymmetrical 180\degr\ and fully symmetrical 90\degr\ hybrids.
Both have 180\degr\ phase total shifts between the outputs, but the
shifts are distributed differently relative to the input signals.

Computing the multiplier output $v_{out}$ is easiest when the signals
are in complex phasor notation with implicit time dependence, $v(t)
\equiv |v|e^{j\xi}$.  Then the low frequency correlator output is
$v_{out} \propto \langle {\rm Re} (v_A v_B^*) \rangle$ where $v_{A,B}$
are voltages at the multiplier input and the asterisk denotes the
complex conjugate.

The most useful case has gain and noise following the hybrid.  Solving
for the multiplier's output shows that the interesting correlator
power-difference signal $v_{out} \propto (\langle x^2 \rangle - \langle
y^2 \rangle)$ is largest when a term
$\cos(\zeta_A-\zeta_B-\varphi+\theta)$ is maximum.  Here $\theta$ is
the hybrid phase defined in Eq.~(\ref{eq:smatrix}), $\varphi$ is an
additional system phase shift shown in Fig.~\ref{fig:blkcrx}, and
$\zeta_{A,B}$ are the gain phases $g_A g_B^* \equiv G e^{j
(\zeta_A-\zeta_B)}$.  Allowing for a phase deviation $\delta$, ideally
zero, from the maximum difference condition, this cosine term is
maximum for $\varphi = \theta + \zeta_A - \zeta_B - \delta$.  With
that substitution the expressions become much simpler, and the
multiplier output is
\begin{widetext}
\begin{eqnarray}
v_{out} &\propto& \Big[ \alpha\beta \left( \langle x^2 \rangle - \langle
  y^2  \rangle \right) 
+ \left( \beta^2  - \alpha^2 \right) \langle xy \rangle
+ \frac{1}{2}\Big( \langle n_{AI}n_{BI} \rangle + \langle n_{AQ}n_{BQ} 
\rangle \Big)
\nonumber\\  
&& + \frac{\alpha}{\sqrt{2}} \Big( \langle xn_{AI} \rangle - \langle yn_{BI}
\rangle \Big)
+ \frac{\beta}{\sqrt{2}} \Big(\langle xn_{BI} \rangle + \langle yn_{AI}
\rangle \Big) 
 \Big] G \cos(\delta)
\nonumber\\  
&+& \Big[ 
\frac{1}{2}\Big( \langle n_{AQ}n_{BI} \rangle - \langle n_{AI}n_{BQ} 
\rangle \Big)
+ \frac{\alpha}{\sqrt{2}} \Big( \langle xn_{AQ} \rangle + \langle yn_{BQ} 
\rangle \Big)
\nonumber\\ && 
+ \frac{\beta}{\sqrt{2}} \Big( \langle xn_{BQ} \rangle - \langle yn_{AQ}   
\rangle \Big) 
 \Big] G \sin(\delta)
\label{eq:outinph} 
\end{eqnarray}
for a 180\degr\ hybrid ($\theta = 0$), and
\begin{eqnarray}
v_{out} &\propto& \Big[ 
\alpha\beta \left( \langle x^2 \rangle - \langle y^2 \rangle \right) 
+ \frac{1}{2}\Big( \langle n_{AQ}n_{BI} \rangle - \langle n_{AI}n_{BQ} 
\rangle \Big)
\nonumber\\ 
&& + \frac{\alpha}{\sqrt{2}} \Big( \langle xn_{AI} \rangle - \langle yn_{BI} 
\rangle \Big) 
- \frac{\beta}{\sqrt{2}} \Big( \langle xn_{BQ} \rangle - \langle yn_{AQ}   
\rangle \Big)
 \Big] G \cos(\delta) 
\nonumber\\ 
&-& \Big[ 
\left( \beta^2 + \alpha^2 \right) \langle xy \rangle
+ \frac{1}{2} \Big( \langle n_{AI}n_{BI} \rangle + \langle n_{AQ}n_{BQ}
\rangle \Big) 
\nonumber\\ &&
+ \frac{\alpha}{\sqrt{2}} \Big( \langle xn_{AQ} \rangle + \langle yn_{BQ} 
\rangle \Big)
+ \frac{\beta}{\sqrt{2}} \Big( \langle xn_{BI} \rangle + \langle yn_{AI} 
\rangle \Big)
\Big] G \sin(\delta) 
\label{eq:outquad}
\end{eqnarray}
for a 90\degr\ hybrid ($\theta = -\pi/2$).
\end{widetext}

Ideally, the correlator output contains only the power-difference term
$\langle x^2 \rangle - \langle y^2 \rangle$.  Other uncorrelated
cross-terms (e.g.\ $\langle x n_A \rangle$) average to zero as
$1/\sqrt{B \tau}$.  Not all elements within the cross-terms will be
completely uncorrelated in a practical system, with the correlated
portions producing offsets at the correlator output.  Fluctuations in
system gain $G$ (Eqs.~\ref{eq:outinph} and \ref{eq:outquad}) scale the
offsets and can produce error terms that are large compared with the
signal.  An important goal for a continuous comparison radiometer is
to keep the multiplier output near zero so the influence of gain
fluctuations will be small.  Minimizing the number and amplitude of
correlator offsets is important to this end.

For ground-based radio astronomy the most important difference between
the circuits with 180\degr\ and 90\degr\ hybrids is likely to be the
response to cross-correlation in the input signals, $\langle xy
\rangle$.  Atmospheric emission in the telescope's near field and
noise from the telescope's ohmic loss will provide correlated voltages
between the two inputs.  Suppressing this term requires either good
hybrid amplitude balance or a flat system phase: at the correlator
output this signal scales with $(\alpha^2 - \beta^2)$ for the
180\degr\ hybrid and $\sim \sin(\delta)$ for the 90\degr\ hybrid.  A
factor of ten suppression implies a hybrid power imbalance no worse
than 0.45/0.55 (0.87 dB) or a maximum phase error term of
$\sin(5.7^\circ)$.  Although it is difficult to build hybrids with
tight amplitude matching across wide bands, wideband hybrids can have
good phase flatness \cite{srikanth01, kooi04}.  If the other system
components have good phase matching then the 90\degr\ hybrid would be
the better choice.

Rejecting correlated input noise can also be useful when a common
local oscillator signal (LO) is injected before the hybrid.  Injecting
the LO into both ports with zero phase shift and equal amplitude will
suppress noise in the oscillator's wings at the signal frequency.
Injecting the LO into only one input port will pump both mixers but
provides no LO noise rejection. 

In any case, the 90\degr\ hybrid circuit always rejects correlated
signals introduced after the hybrid better than the 180\degr\ hybrid
circuit.  These terms, with form $\langle n_{AQ} n_{BQ}\rangle$ and
$\langle n_{AI} n_{BI}\rangle$, are suppressed by the $\sin(\delta)$
factor.  Correlation in these terms can arise from bias fluctuations
common to both gain paths or noise from the wings of a shared local
oscillator (the noise and gain model implicit in Fig.~\ref{fig:blkcrx}
is generic and can include frequency conversion and multiple
amplifiers).

For a nonideal hybrid, coupling between the output ports (2 and 3 in
Fig.~\ref{fig:crxblock}) is another important potential source of
correlator offset.  Noise radiated from, or signals reflected by,
devices following the hybrid can emerge from the corresponding
nominally isolated port to produce a correlated signal.  A circulator
following the hybrid can reduce the offset by an amount equal to the
circulator isolation at the cost of adding loss before amplification.
Lack of isolation between hybrid inputs is likely to be less of a
problem since the signal reflected from the telescope or other optics
is likely to be small.  Further weak correlated terms will come from
noise power emitted from the system and reflected back as an input
signal (e.g.\ a fraction of $n_A$ returns as input signal $x$ to
produce a nonzero $\langle x n_A \rangle$).  This term is suppressed
if the pathlength for the reflected signal is substantially longer
than the correlation length, $l \simeq c/B$, where $c$ is the
speed of light.

Phase switching is a powerful method for removing correlator offsets
and reducing the effects from nonideal colored noise.  In comparison
with amplitude modulation (Dicke switching) 180\degr\ phase switching
is very efficient because the full signal amplitude is always present
at the detector.  Modulating the phase difference between the arms of
a continuous comparison radiometer shifts the correlated signal output
in frequency by an amount equal to the modulation frequency.
Synchronous detection recovers the correlated signal while rejecting
noise fluctuations at frequencies other than the modulation frequency.
With phase switching before amplification, ideally just following the
hybrid, a judicious choice of modulation frequency can remove much of
the drift and 1/$f$ noise associated with amplifier noise
fluctuations.  Since, to high order, any residual offsets or offsets
from sources outside the phase modulation-demodulation boundaries are
independent of the telescope's pointing, Dicke amplitude modulation by
chopping on the sky (optically switching between two positions on the
sky) and a final synchronous demodulation will largely remove the
remaining offsets.

Phase switching does not suppress the direct effects of amplifier or
multiplier gain or phase fluctuations, but symmetry can reduce their
influence.  Equations~(\ref{eq:outinph}) and (\ref{eq:outquad}) show
that the product of the voltage gains scales the input power
difference as $G( \langle x^2 \rangle - \langle y^2 \rangle )$.  A
small gain fluctuation common to both chains introduces an amplitude
error to the difference signal at the correlator output, but adds no
error signal when the offsets are negligible.  Multiplier gain
fluctuations have the same effect as amplifier gain fluctuations in
this case. The situation is different when the amplifier gain
fluctuations are differential-mode instead of common-mode for the two
amplifier chains.  Faris \cite{faris67} calculates the effect of a
varying gain imbalance between the two chains for $g_2(f,t) = [1 +
a(t)]g_1(f)$, where $g_{1,2}(f)$ are the complex amplifier voltage
gains of the two chains and $a(t)$ is a zero-mean random variable that
describes the differential-mode fluctuations.  Fluctuations increase
the output variance by a factor of $(1 + \langle a^2(t) \rangle)$
compared with the case of purely common-mode gain fluctuations between
the two arms.  Similar effects arise from differential phase
fluctuations between the two arms.  Strategies for minimizing
differential-mode fluctuations could include biasing the amplifiers
from a common power supply and keeping good thermal contact between
corresponding parts of each chain.

The second case, with amplification preceding the hybrid's loss, is
the obvious choice for maximizing the receiver sensitivity but negates
much of the continuous comparison architecture's advantage.  The maximum
difference signal in this case is
\begin{eqnarray}
v_{out} &=& \alpha\beta \Big[ 
\left( \langle x^2 \rangle |g_x|^2 - \langle y^2 \rangle |g_y|^2
\right) 
\nonumber \\&&
+ \left( \langle n_x^2 \rangle |g_x|^2 
- \langle n_y^2 \rangle |g_y|^2 \right)
\Big] \; ,
\label{eq:upfront}
\end{eqnarray}
for both 180\degr\ and 90\degr\ hybrids.  For this configuration to be
useful nearly exact matching of both the noise power and power gains
would be necessary. Slight imbalances in loss and gain will produce
large offsets at the correlator output.

\section{Spectroscopy with an analog lag cross-correlator \label{sec:anaxc}} 

Sensitive spectroscopy (multichannel radiometry) over broad
bandwidths is important for observations of wide spectral lines from
distant galaxies, searches for lines at with uncertain frequency, and
measurements of pressure-broadened lines in planetary atmospheres.
Spectrometer bandwidths can be tens of gigahertz with channel
bandwidths of tens of megahertz.  Such broad bandwidths place
stringent requirements on system stability, so it is natural to pair
wideband spectrometers with the continuous comparison architecture.
Spectroscopy with a correlation radiometer requires measurement of the
cross-correlation function over a range of time lag, or delay.  Analog
lag correlators use purely analog components to obtain the
cross-correlation function $R_{AB}(\tau)$ as a function of lag $\tau$:
\begin{equation}
R_{AB}(\tau) =  \lim_{T\to\infty} \frac{1}{2T} \int^T_{-T} v_A(t) \cdot
v_B(t+\tau) \, \mathrm{d}t \; .
\end{equation}
Tapped transmission lines provide the time delays $\tau$, transistor
multipliers form the product of the two input voltages $v_A(t)$ and
$v_B(t+\tau)$, and low-frequency electronics integrate the multiplier
output to provide the time average.  

A Fourier transform of the cross-correlation function yields the power
density cross-spectrum.  Transforming the correlation function to
recover the spectrum is slightly more complicated than making a direct
Fourier transform for analog correlators because the signal is not
sampled at perfectly regular intervals. Although the mechanical
spacing between the microwave signal taps along the transmission line
is well defined by the traces on the circuit board,
frequency-dependent component variations cause small erratic
variations in the electrical delays between multipliers.  A direct
Fourier inversion to find the spectrum is not possible because the
transform kernel's phase term cannot be reduced to a separable product
of the delay time and frequency, as might be the case for simple line
dispersion \cite{hz01}. It is possible to correct the irregular
sampling in software by measuring the spectrometer's response to a
series of monochromatic signals at known frequencies, then expanding
the astronomical input signal on these measured functions
\cite{hz01}.

An interesting effect of this calibration scheme is that it defines
the spectrometer's internal phase: by definition signals in phase with
the calibration are real, and those with a relative 90\degr\ phase
shift are imaginary.  This property can be used to eliminate one of
the phases in the usual complex cross-correlation measurement.  When
the calibration signals are injected at the radiometer's input the
calibration and measured signals share the same phase shifts through
the entire instrument, so the measured signal is purely real; the
imaginary component contains only noise.  A single cross-correlator
can therefore measure the cross-correlation function.  While purely
real cross-correlations are unusual in most spectrometers, there is no
fundamental reason that they cannot exist.  A real correlation
function has the convenient property of even symmetry in the lag
domain, so the positive (or negative) lags alone contain all the
necessary information to recover the spectrum.  Although it is
possible to build a full complex correlator and calibrate it at lower
frequency, injecting the phase calibration signals at the input to the
entire radiometer yields spectra at full spectral resolution with half
the number of lags.  This is not only a significant savings in
spectrometer cost and complexity, but eliminates requirements on
phase-matching between the two amplifier and processing chains.
Digital correlators do not readily share this property because their
topologies give them an intrinsic symmetry and phase related to the
position of the zero-lag multiplier.  A full complex correlator, or
close phase matching across the receiver band, is needed for
spectroscopy with a digital system and direct transform.

\begin{figure}
\includegraphics[width=2.5in]{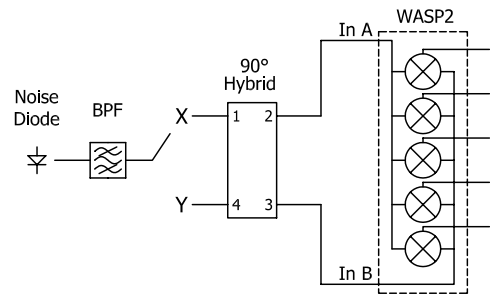} 
\caption{Block diagram of laboratory continuous comparison receiver test
  setup. \label{fig:crxblock}}
\end{figure}

\begin{figure}
\includegraphics[width=3.2in]{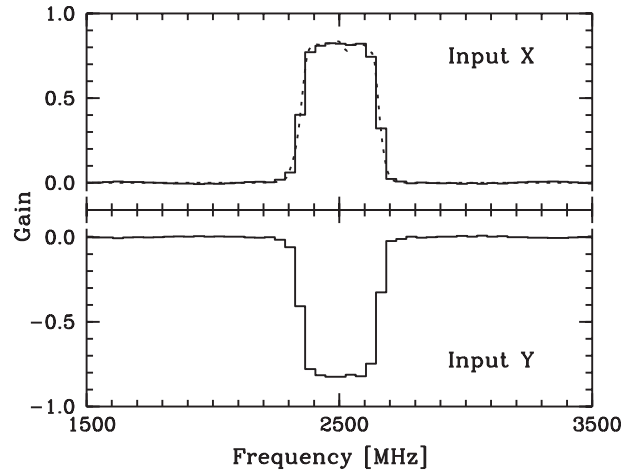} 
\caption{Spectra from the two inputs of a laboratory correlation
radiometer with a WASP2 spectrometer configured as a cross-correlator.
The dotted line is a network analyzer measurement of the filter
transmission.
\label{fig:xcspec}}
\end{figure}

A simple laboratory continuous comparison radiometer, sketched in
Figure~\ref{fig:crxblock}, verifies that a WASP2 (Wideband
Astronomical SPectrometer) analog lag correlator \cite{hz01} properly
produces power difference cross-spectra with this calibration scheme.
The hybrid for the experiment was an off-the-shelf stripline 90\degr\
2--4~GHz device.  Cables lengths between the hybrid and
cross-correlator brought the zero time-lag position close to one end
of the multiplier ladder, for maximum spectral resolution, but were
not otherwise trimmed for length or phase matching. A broadband noise
diode and 300~MHz filter generated an artificial spectral line that
could be connected to either hybrid input, denoted X and Y in
Fig.~\ref{fig:crxblock}.  Phase calibration signals were fed into
input X alone.  Figure~\ref{fig:xcspec} shows that the spectrometer
works as predicted.  The artificial line at input X, with Y
terminated, produces a positive spectral line, and the same signal at
input Y, with X terminated, produces a negative spectral line.  After
removing the noise source's intrinsic spectral shape by dividing the
raw spectra by spectra of the noise source alone, the filter center
frequency, shape, and loss matched network analyzer measurements
(Fig.~\ref{fig:xcspec}).  This confirms that the cross-correlation
function is purely real.  The sum of the two spectra is very close to
zero.  Residual phase errors scatter power across the spectrum at
about 0.5\% of the peak line intensity, a well-understood dynamic
range limit rather than a noise offset \cite{hz01}.  This
fixed-pattern structure subtracts well with beamswitching.

These experimental results show that the combination of the continuous
comparison architecture with a analog lag correlator is a very
promising method for spectroscopy over very wide bandwidths.

\begin{acknowledgments}

The author thanks J.~Kooi and T.G.~Phillips for suggestions and
several useful discussions about LO noise rejection and practical
hybrid designs.  The author also thanks the Caltech Submillimeter
Group for their hospitality while this paper was written.  This work
was supported in part by grant AST-9819747 from the National Science
Foundation.

\end{acknowledgments}


\begin{thebibliography}{22}
\expandafter\ifx\csname natexlab\endcsname\relax\def\natexlab#1{#1}\fi
\expandafter\ifx\csname bibnamefont\endcsname\relax
  \def\bibnamefont#1{#1}\fi
\expandafter\ifx\csname bibfnamefont\endcsname\relax
  \def\bibfnamefont#1{#1}\fi
\expandafter\ifx\csname citenamefont\endcsname\relax
  \def\citenamefont#1{#1}\fi
\expandafter\ifx\csname url\endcsname\relax
  \def\url#1{\texttt{#1}}\fi
\expandafter\ifx\csname urlprefix\endcsname\relax\def\urlprefix{URL }\fi
\providecommand{\bibinfo}[2]{#2}
\providecommand{\eprint}[2][]{\url{#2}}

\bibitem[{\citenamefont{Harris and Zmuidzinas}(2001)}]{hz01}
\bibinfo{author}{\bibfnamefont{A.}~\bibnamefont{Harris}} \bibnamefont{and}
  \bibinfo{author}{\bibfnamefont{J.}~\bibnamefont{Zmuidzinas}},
  \bibinfo{journal}{Rev. Sci. Inst.} \textbf{\bibinfo{volume}{72}},
  \bibinfo{pages}{1531} (\bibinfo{year}{2001}).

\bibitem[{\citenamefont{Haslam et~al.}(1974)\citenamefont{Haslam, Wilson,
  Graham, and Hunt}}]{haslam74}
\bibinfo{author}{\bibfnamefont{C.}~\bibnamefont{Haslam}},
  \bibinfo{author}{\bibfnamefont{W.}~\bibnamefont{Wilson}},
  \bibinfo{author}{\bibfnamefont{D.}~\bibnamefont{Graham}}, \bibnamefont{and}
  \bibinfo{author}{\bibfnamefont{G.}~\bibnamefont{Hunt}},
  \bibinfo{journal}{Astron. Astrophys. Suppl.} \textbf{\bibinfo{volume}{13}},
  \bibinfo{pages}{359} (\bibinfo{year}{1974}).

\bibitem[{\citenamefont{Jarosik et~al.}(2003)\citenamefont{Jarosik, Bennet,
  Halpern, Hinshaw, Kogut, Limon, Meyer, Page, Pospiezalski, Spergel
  et~al.}}]{jarosik03a}
\bibinfo{author}{\bibfnamefont{N.}~\bibnamefont{Jarosik}},
  \bibinfo{author}{\bibfnamefont{C.}~\bibnamefont{Bennet}},
  \bibinfo{author}{\bibfnamefont{M.}~\bibnamefont{Halpern}},
  \bibinfo{author}{\bibfnamefont{G.}~\bibnamefont{Hinshaw}},
  \bibinfo{author}{\bibfnamefont{A.}~\bibnamefont{Kogut}},
  \bibinfo{author}{\bibfnamefont{M.}~\bibnamefont{Limon}},
  \bibinfo{author}{\bibfnamefont{S.}~\bibnamefont{Meyer}},
  \bibinfo{author}{\bibfnamefont{L.}~\bibnamefont{Page}},
  \bibinfo{author}{\bibfnamefont{M.}~\bibnamefont{Pospiezalski}},
  \bibinfo{author}{\bibfnamefont{D.}~\bibnamefont{Spergel}},
  \bibnamefont{et~al.}, \bibinfo{journal}{Astrophys. J. Suppl. Ser.}
  \textbf{\bibinfo{volume}{145}}, \bibinfo{pages}{413} (\bibinfo{year}{2003}).

\bibitem[{\citenamefont{Predmore et~al.}(1985)\citenamefont{Predmore, Erickson,
  Huguenin, and Goldsmith}}]{predmore85}
\bibinfo{author}{\bibfnamefont{C.}~\bibnamefont{Predmore}},
  \bibinfo{author}{\bibfnamefont{N.}~\bibnamefont{Erickson}},
  \bibinfo{author}{\bibfnamefont{G.}~\bibnamefont{Huguenin}}, \bibnamefont{and}
  \bibinfo{author}{\bibfnamefont{P.}~\bibnamefont{Goldsmith}},
  \bibinfo{journal}{IEEE Trans. Microwave Theory Tech.}
  \textbf{\bibinfo{volume}{MTT-33}}, \bibinfo{pages}{44}
  (\bibinfo{year}{1985}).

\bibitem[{\citenamefont{Rholfs and Wilson}(2000)}]{rholfs00_crx}
\bibinfo{author}{\bibfnamefont{K.}~\bibnamefont{Rholfs}} \bibnamefont{and}
  \bibinfo{author}{\bibfnamefont{T.}~\bibnamefont{Wilson}},
  \emph{\bibinfo{title}{Tools of Radio Astronomy}}
  (\bibinfo{publisher}{Springer Verlag}, \bibinfo{year}{2000}), chap.
  \bibinfo{chapter}{4.5}, \bibinfo{edition}{3rd} ed.

\bibitem[{\citenamefont{Koistinen et~al.}(2002)\citenamefont{Koistinen,
  Lahtinen, and Hallikainen}}]{koistinen02}
\bibinfo{author}{\bibfnamefont{O.}~\bibnamefont{Koistinen}},
  \bibinfo{author}{\bibfnamefont{J.}~\bibnamefont{Lahtinen}}, \bibnamefont{and}
  \bibinfo{author}{\bibfnamefont{M.~T.} \bibnamefont{Hallikainen}},
  \bibinfo{journal}{IEEE Trans.\ Instrumen.\ Meas.}
  \textbf{\bibinfo{volume}{IM-51}}, \bibinfo{pages}{227}
  (\bibinfo{year}{2002}).

\bibitem[{\citenamefont{Fano}(1951)}]{fano51}
\bibinfo{author}{\bibfnamefont{R.}~\bibnamefont{Fano}},
  \emph{\bibinfo{title}{Signal to noise ratio in correlation detectors}},
  \bibinfo{howpublished}{M.I.T. Res. Lab. Elec., Rep.\ RLE-186}
  (\bibinfo{year}{1951}).

\bibitem[{\citenamefont{Goldstein}(1955)}]{gs55a}
\bibinfo{author}{\bibfnamefont{S.}~\bibnamefont{Goldstein},
  \bibfnamefont{Jr.}}, \bibinfo{journal}{Proc. IRE}
  \textbf{\bibinfo{volume}{43}}, \bibinfo{pages}{1663} (\bibinfo{year}{1955}),
  \bibinfo{note}{with further correspondence and errata in Proc. IRE {\bf 48},
  365, 1956}.

\bibitem[{\citenamefont{Blum}(1959)}]{blum59}
\bibinfo{author}{\bibfnamefont{{\'E}.-J.} \bibnamefont{Blum}},
  \bibinfo{journal}{Ann. Astrophys.} \textbf{\bibinfo{volume}{22}},
  \bibinfo{pages}{140} (\bibinfo{year}{1959}).

\bibitem[{\citenamefont{Tiuri}(1964)}]{tiuri64}
\bibinfo{author}{\bibfnamefont{M.}~\bibnamefont{Tiuri}}, \bibinfo{journal}{IEEE
  Trans. Antenn. Propag.} \textbf{\bibinfo{volume}{AP-12}},
  \bibinfo{pages}{930} (\bibinfo{year}{1964}).

\bibitem[{\citenamefont{Colvin}(1961)}]{colvin61}
\bibinfo{author}{\bibfnamefont{R.}~\bibnamefont{Colvin}}, Ph.D. thesis,
  \bibinfo{school}{Stanford University} (\bibinfo{year}{1961}),
  \bibinfo{note}{also Stanford Radio Astronomy Institute Publ.\ No.\ 18A}.

\bibitem[{\citenamefont{Faris}(1967)}]{faris67}
\bibinfo{author}{\bibfnamefont{J.}~\bibnamefont{Faris}}, \bibinfo{journal}{J.
  Res. Nat. Bur. Stand.--C} \textbf{\bibinfo{volume}{71C}},
  \bibinfo{pages}{153} (\bibinfo{year}{1967}).

\bibitem[{\citenamefont{{Seiffert} et~al.}(2002)\citenamefont{{Seiffert},
  {Mennella}, {Burigana}, {Mandolesi}, {Bersanelli}, {Meinhold}, and
  {Lubin}}}]{seiffert02}
\bibinfo{author}{\bibfnamefont{M.}~\bibnamefont{{Seiffert}}},
  \bibinfo{author}{\bibfnamefont{A.}~\bibnamefont{{Mennella}}},
  \bibinfo{author}{\bibfnamefont{C.}~\bibnamefont{{Burigana}}},
  \bibinfo{author}{\bibfnamefont{N.}~\bibnamefont{{Mandolesi}}},
  \bibinfo{author}{\bibfnamefont{M.}~\bibnamefont{{Bersanelli}}},
  \bibinfo{author}{\bibfnamefont{P.}~\bibnamefont{{Meinhold}}},
  \bibnamefont{and} \bibinfo{author}{\bibfnamefont{P.}~\bibnamefont{{Lubin}}},
  \bibinfo{journal}{Astr. Astrophys.} \textbf{\bibinfo{volume}{391}},
  \bibinfo{pages}{1185} (\bibinfo{year}{2002}).

\bibitem[{\citenamefont{{Mennella} et~al.}(2003)\citenamefont{{Mennella},
  {Bersanelli}, {Seiffert}, {Kettle}, {Roddis}, {Wilkinson}, and
  {Meinhold}}}]{mennella03}
\bibinfo{author}{\bibfnamefont{A.}~\bibnamefont{{Mennella}}},
  \bibinfo{author}{\bibfnamefont{M.}~\bibnamefont{{Bersanelli}}},
  \bibinfo{author}{\bibfnamefont{M.}~\bibnamefont{{Seiffert}}},
  \bibinfo{author}{\bibfnamefont{D.}~\bibnamefont{{Kettle}}},
  \bibinfo{author}{\bibfnamefont{N.}~\bibnamefont{{Roddis}}},
  \bibinfo{author}{\bibfnamefont{A.}~\bibnamefont{{Wilkinson}}},
  \bibnamefont{and}
  \bibinfo{author}{\bibfnamefont{P.}~\bibnamefont{{Meinhold}}},
  \bibinfo{journal}{Astr. Astrophys.} \textbf{\bibinfo{volume}{410}},
  \bibinfo{pages}{1089} (\bibinfo{year}{2003}).

\bibitem[{\citenamefont{Ryle}(1952)}]{ryle52}
\bibinfo{author}{\bibfnamefont{M.}~\bibnamefont{Ryle}}, \bibinfo{journal}{Proc.
  R. Soc. Lon. A} \textbf{\bibinfo{volume}{211}}, \bibinfo{pages}{351}
  (\bibinfo{year}{1952}).

\bibitem[{\citenamefont{Lee et~al.}(1949)\citenamefont{Lee, Chetham, and
  Wiesner}}]{lee49}
\bibinfo{author}{\bibfnamefont{Y.}~\bibnamefont{Lee}},
  \bibinfo{author}{\bibfnamefont{T.}~\bibnamefont{Chetham}}, \bibnamefont{and}
  \bibinfo{author}{\bibfnamefont{J.}~\bibnamefont{Wiesner}},
  \emph{\bibinfo{title}{The application of correlation functions in the
  detections of small signals in noise}}, \bibinfo{howpublished}{M.I.T. Res.
  Lab. Elec., Rep.\ RLE-141} (\bibinfo{year}{1949}).

\bibitem[{\citenamefont{Lee and Wiesner}(1950)}]{lee50}
\bibinfo{author}{\bibfnamefont{Y.}~\bibnamefont{Lee}} \bibnamefont{and}
  \bibinfo{author}{\bibfnamefont{J.}~\bibnamefont{Wiesner}},
  \bibinfo{journal}{Electronics} \textbf{\bibinfo{volume}{23}},
  \bibinfo{pages}{86} (\bibinfo{year}{1950}).

\bibitem[{\citenamefont{Peebles}(1993)}]{pzpeebles93_corr}
\bibinfo{author}{\bibfnamefont{P.}~\bibnamefont{Peebles}},
  \emph{\bibinfo{title}{Probability, Random Variables, and Random Signal
  Processes}} (\bibinfo{publisher}{McGraw-Hill}, \bibinfo{year}{1993}), p.
  \bibinfo{pages}{310}, \bibinfo{edition}{3rd} ed.

\bibitem[{\citenamefont{Dicke}(1946)}]{dicke46}
\bibinfo{author}{\bibfnamefont{R.}~\bibnamefont{Dicke}}, \bibinfo{journal}{Rev.
  Sci. Inst.} \textbf{\bibinfo{volume}{17}}, \bibinfo{pages}{268}
  (\bibinfo{year}{1946}).

\bibitem[{\citenamefont{Srikanth and Kerr}(2001)}]{srikanth01}
\bibinfo{author}{\bibfnamefont{S.}~\bibnamefont{Srikanth}} \bibnamefont{and}
  \bibinfo{author}{\bibfnamefont{A.}~\bibnamefont{Kerr}},
  \emph{\bibinfo{title}{Waveguide quadrature hybrids for {ALMA} receivers}},
  \bibinfo{howpublished}{ALMA Memo 343} (\bibinfo{year}{2001}),
  \urlprefix\url{http://www.alma.nrao.edu/memos/index.html}.

\bibitem[{\citenamefont{Kooi et~al.}(2004)\citenamefont{Kooi, Kovacs, Bumble,
  Ghattopadhyay, Edgar, Kaye, LeDuc, Zmuidzinas, and Phillips}}]{kooi04}
\bibinfo{author}{\bibfnamefont{J.}~\bibnamefont{Kooi}},
  \bibinfo{author}{\bibfnamefont{A.}~\bibnamefont{Kovacs}},
  \bibinfo{author}{\bibfnamefont{B.}~\bibnamefont{Bumble}},
  \bibinfo{author}{\bibfnamefont{G.}~\bibnamefont{Ghattopadhyay}},
  \bibinfo{author}{\bibfnamefont{M.}~\bibnamefont{Edgar}},
  \bibinfo{author}{\bibfnamefont{S.}~\bibnamefont{Kaye}},
  \bibinfo{author}{\bibfnamefont{H.}~\bibnamefont{LeDuc}},
  \bibinfo{author}{\bibfnamefont{J.}~\bibnamefont{Zmuidzinas}},
  \bibnamefont{and} \bibinfo{author}{\bibfnamefont{T.}~\bibnamefont{Phillips}},
  in \emph{\bibinfo{booktitle}{Millimeter and submillimeter detectors for
  astronomy II}}, edited by
  \bibinfo{editor}{\bibfnamefont{J.}~\bibnamefont{Zmuidzinas}},
  \bibinfo{editor}{\bibfnamefont{W.~S.} \bibnamefont{Holland}},
  \bibnamefont{and}
  \bibinfo{editor}{\bibfnamefont{S.}~\bibnamefont{Withington}}
  (\bibinfo{publisher}{SPIE--The International Society for Optical
  Engineering}, \bibinfo{year}{2004}), vol. \bibinfo{volume}{4587} of
  \emph{\bibinfo{series}{Proc. SPIE}}, pp. \bibinfo{pages}{332--348}.

\bibitem[{\citenamefont{Pozar}(1998)}]{pozar98_spar}
\bibinfo{author}{\bibfnamefont{D.~M.} \bibnamefont{Pozar}},
  \emph{\bibinfo{title}{Microwave Engineering}} (\bibinfo{publisher}{J.W.\
  Wiley \& Sons}, \bibinfo{year}{1998}), pp. \bibinfo{pages}{354--357},
  \bibinfo{edition}{2nd} ed.

\end{thebibliography}

\end{document}